\documentclass[12pt]{article}
\usepackage{graphicx}
\usepackage[english]{babel}

\begin{document}

\begin{center}
{\large{\bf CUMULATIVE PROCESSES}}

\vspace*{5mm}

\underline{S.S. Shimanskiy}

\footnote{E--mail: {\tt shimansk@sunhe.jinr.ru}}

\vspace*{3mm}

{\it JINR, 141980, Dubna, Moscow region, Russia} \\
\end{center}

\vspace*{5mm}

{\small{ \centerline{\bf Abstract} In this report we talk about
nowadays situation with the cumulative effect studies. The
experimental programm for the nuclotron and other facilities is
proposed. }} \vspace*{3mm}

{\bf 1.~~Introduction}

\vspace*{3mm}

The simple picture that nuclei are consisted of nucleons have only
limit range of validity. We know some number of intrigue
phenomenons which have no adequate understanding in frame of the
naive nucleon model of nuclei. These phenomenons are connected
with states of the nuclear matter when nucleons must be overlapped
inside the nuclei and the baryon matter converts to the
quark-gluon phase of nuclear matter. Through quantum nature the
short distances mean the high transfer momentum processes or
states under the very high pressure(as we can wait in some stages
of the star evolution). Long time investigations of high transfer
momentum processes with nuclei give us to say about some unified
explanations for their nature. Nowadays we saw great growing
interest to the astrophysics problems. We could not know stars
evolution laws without understanding of the short distance
nucleon-nucleon interaction. Cumulative phenomenons shows that we
can wait formation of very exotic star states which haven't
investigate before this time(for examples, baryon stars or
bosoning stars).

Historically the first of these phenomenons were deep subthreshold
particle production processes. Second ones were processes of
direct knock out of deuterons in pA-interactions with the very
high probability. The birth day of the cumulative processes was
1971 when it had been predicted by A.M. Baldin [1] and short time
later was discovered by team of V.S. Stavinsky [2]. In 1993 A. A.
Baldin showed [3] that using scaling variables proposed by V. S.
Stavinsky [4] there is possibility to describe data on cumulative
and subthreshold particle production processes in a same way. This
means that the whole set of phenomenons on nuclei in the
kinematical region far from the kinematical threshold for free
nucleon processes can be considered as processes with the same
nature. We can propose to name all these phenomenons as cumulative
processes. Because this name(from my point of view) can more
adequate reflects their common collective nature.

We will try to show that to isolate the production mechanisms of
cumulative processes will need to measure additional observables
such as associated multiplicities and polarization
characteristics. And the most important that these investigation
must to be carry out in the region of high transverse momentum
with the half-exclusive(and exclusive) setting of experiments.
Moreover the energy region for these studies may be limited by
maximal energies 20-30 Gev for proton beams.

Nuclotron is the accelerator of relativistic nuclei which works
and continues to be improved in the V.I. Veksler and A.M. Baldina
Laboratory of  high energies. The accelerator uses the magnets
with superconducting coils developed in LHE and has been created
to work with proton beams up to energy 12 GeV and  nuclei up to 6
GeV/nucleon. Energies of nuclotron beams temporarily exceed
energies of the synhrophasotron but the main  characteristics of
the new accelerator beams are considerably better and make it
possible to plan experiments which could not be possible to be
carried out earlier. There are some unique nuclotron properties:
the possibility to work with internal beams and have the extracted
beam duration up to 10 seconds. Another special feature of the
accelerating complex lies in the fact that from the middle 1980s
it began the acceleration of polarized deuteron beams. From the
middle 1990s it was mounted and has been working with a polarized
target. Polarization studies are the basic and most important part
of the physical program for nuclotron. The unique cryogenic
targets created by L.B. Golovanov's team are used in LHE. At
present, the possibility to accelerate polarized proton beams [10]
and beams of the polarized $ ^3He $ nuclei on nuclotron is
investigated. In the next 5--7 years nuclotron will not be a
competition.

\vspace*{3mm}

{\bf 2.~~Cumulative processes}

\vspace*{3mm}

The bulk of experimental data of cumulative processes was obtained
by the inclusive setups in fragmentation regions of beams or
targets. Theoretical models have described inclusive spectra of
cumulative particles with different level of agreement but fail
describing all set of the experimental data. Polarization studies
of cumulative processes carried out at LHE [5] resulted in
thorough revision of models, because the data on polarization
characteristics deviated noticeably from first model predictions.
It is still too early to speak about clear understanding of the
nature of cumulative processes. That is why new ideas how to
resolve the puzzle of these phenomenons are so important.

The cumulative data have very nice phenomenological description
using variables introduced by V.S. Stavinsky [4]. In Figure.1
graphically presented definitions Stavinsky's variables. They are
very close to Bjorken's quark-parton model variables.

\begin{center}
\includegraphics[width=150mm,height=120mm]{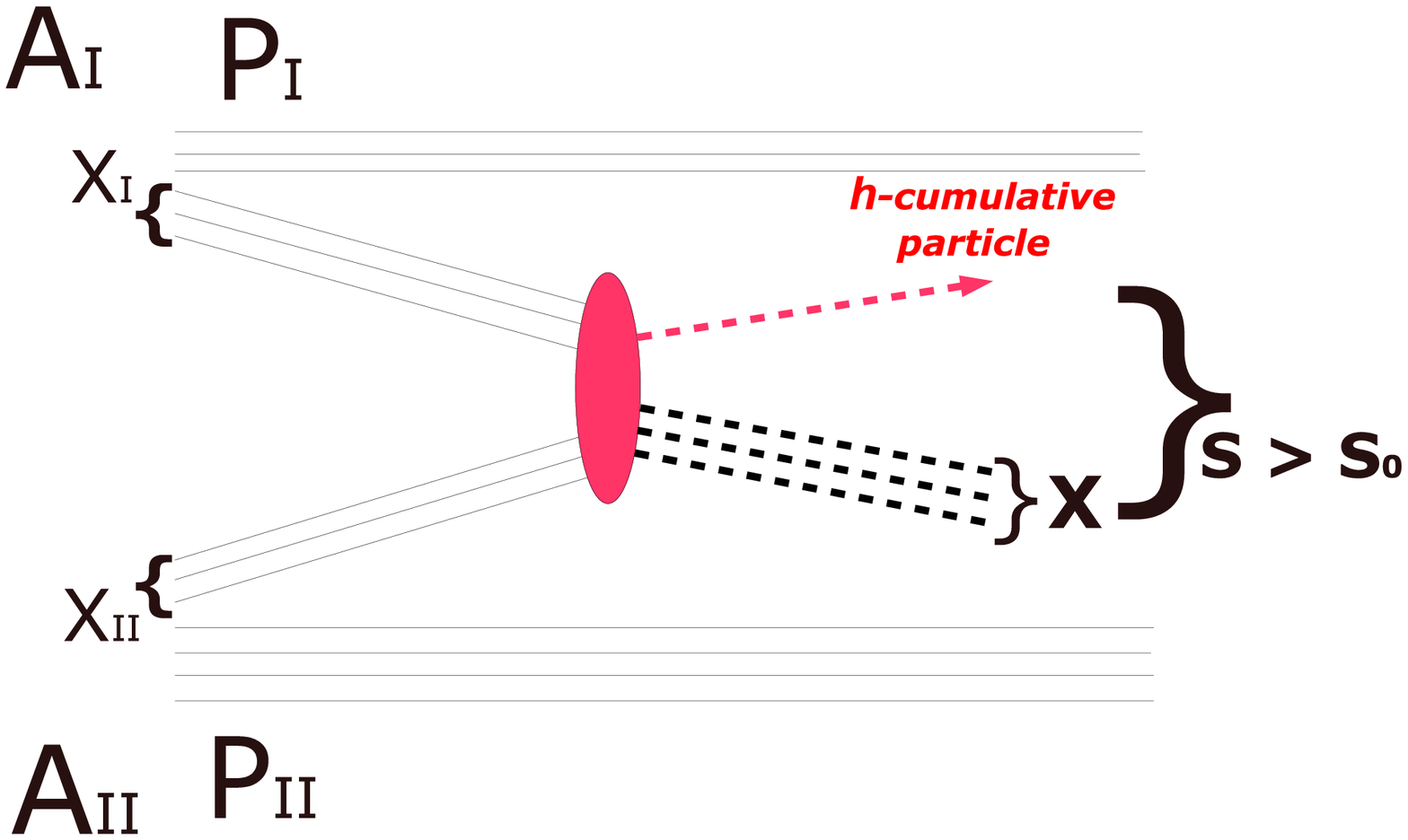} \\
Figure.1 Stavinsky's variables $X_I$ and $X_{II}$\\
\end{center}

In common case Stavinsky's variables describe nuclear-nuclear
collisions of nuclei with atomic weights $A_I$ and $A_{II}$.
Four-momentums for these nuclei are $P_I$ and $P_{II}$.  For
cumulative processes we can subtract subprocesses which
characterized by the invariant value $s_{cumulat}$ which defines
as
$$ {\large s_{cumulat} = (X_{I}\cdot \frac{P_{I}}{A_{I}}
 + X_{II}\cdot \frac{P_{II}}{A_{II}})^2}. $$
The $s_{cumulat}$ need to compare with the invariant value $s_0$
defines as
$$ {\large s_0 = (\frac{P_{I}}{A_{I}}
 + \frac{P_{II}}{A_{II}})^2} $$
which characterized free nucleon-nucleon interactions. Processes
with $s_{cumulat} > s_0$ are forbidding for free nucleon-nucleon
interactions.

The kinematical regions for cumulative processes you can see on
Figure.2.

\includegraphics[width=130mm,height=100mm]{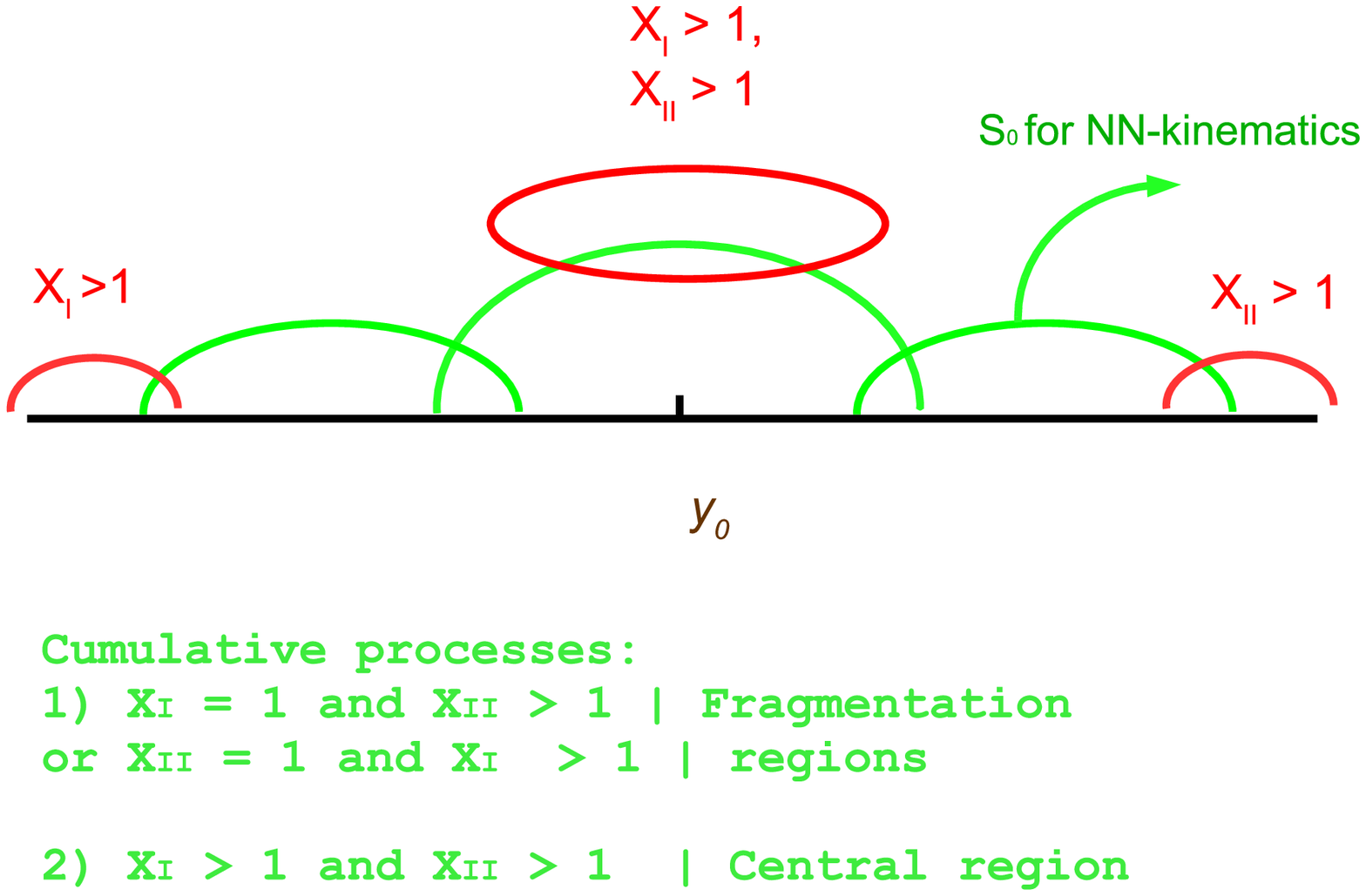} \\
\begin{center}
Figure.2 Cumulative regions Stavinsky's variables $X_I$ and $X_{II}$\\
\end{center}
$y_0$ is a rapidity for the central mass system of nucleon-nucleon
interactions.

Stavinsky's variables have differs compare with $x$ variables for
the quark-parton model. In the quark-parton model $x$ did not fix
and we need to integrate over the full range of $x$. The
additional condition which fix variables have been introduced by
Stavinsky. We need to take $X_I$ and $X_{II}$ only which give us
the minimal value $min(s^{1/2}_{cumulat})$. Combination with the
four-momentum conservation laws give us possibility for unique
determination of variables. Thus defined $X_I$, $X_{II}$  and
$min(s^{1/2}_{cumulat})$ can use for uniform description of
cumulative and subthreshold processes[3] by phenomenological
equation

$$ E\cdot \frac{d^3\sigma}{dp^3}=C_1\cdot A^{\frac{1}{3}+
\frac{X_I}{3}}_I\cdot A^{\frac{1}{3}+\frac{X_{II}}{3}}_{II}\cdot
exp(-\frac{\Pi}{C_2}),$$
where $C_1$ and $C_2$ are constants and
$$ \Pi=\frac{1}{2\cdot m_{nucleon}} \cdot
min(s^{1/2}_{cumulat}),$$
where $m_{nucleon}$ is the nucleon mass.

The nuclei characteristics are  well described when nucleons can
be considered as points and locating in the distances greater than
the sizes of nucleons. Complexities in description of nuclei lie
at the region of small internucleon distances when nucleons in the
nuclei are overlapped and begin to be manifested through
quark-gluon degrees of freedom. It is mean that the nucleon
momentum of Fermi motion must be greater then $\sim 0.3 GeV/c$.
Exactly this is a region of the nonperturbative QCD.

The discovery of the cumulative effect stimulated development of
theoretical models [6] describing reactions with nuclei in a state
in which nucleons strongly overlapped and loose their personality.
The quark-parton model underlies these approaches, since
description of processes with large momentum transfers is
required. All models we can split on two groups. The first ones
say about nucleons with a very high momentum inside nuclei they
are short-range correlation(SRC) models. It is a  some development
of the Fermi motion picture. Main features of SRC are protons with
very high momentums($\bar{k}$) inside nuclei and nucleon-nucleon
interaction vertexes (see Figure.3).

\begin{center}
\includegraphics[width=80mm,height=100mm]{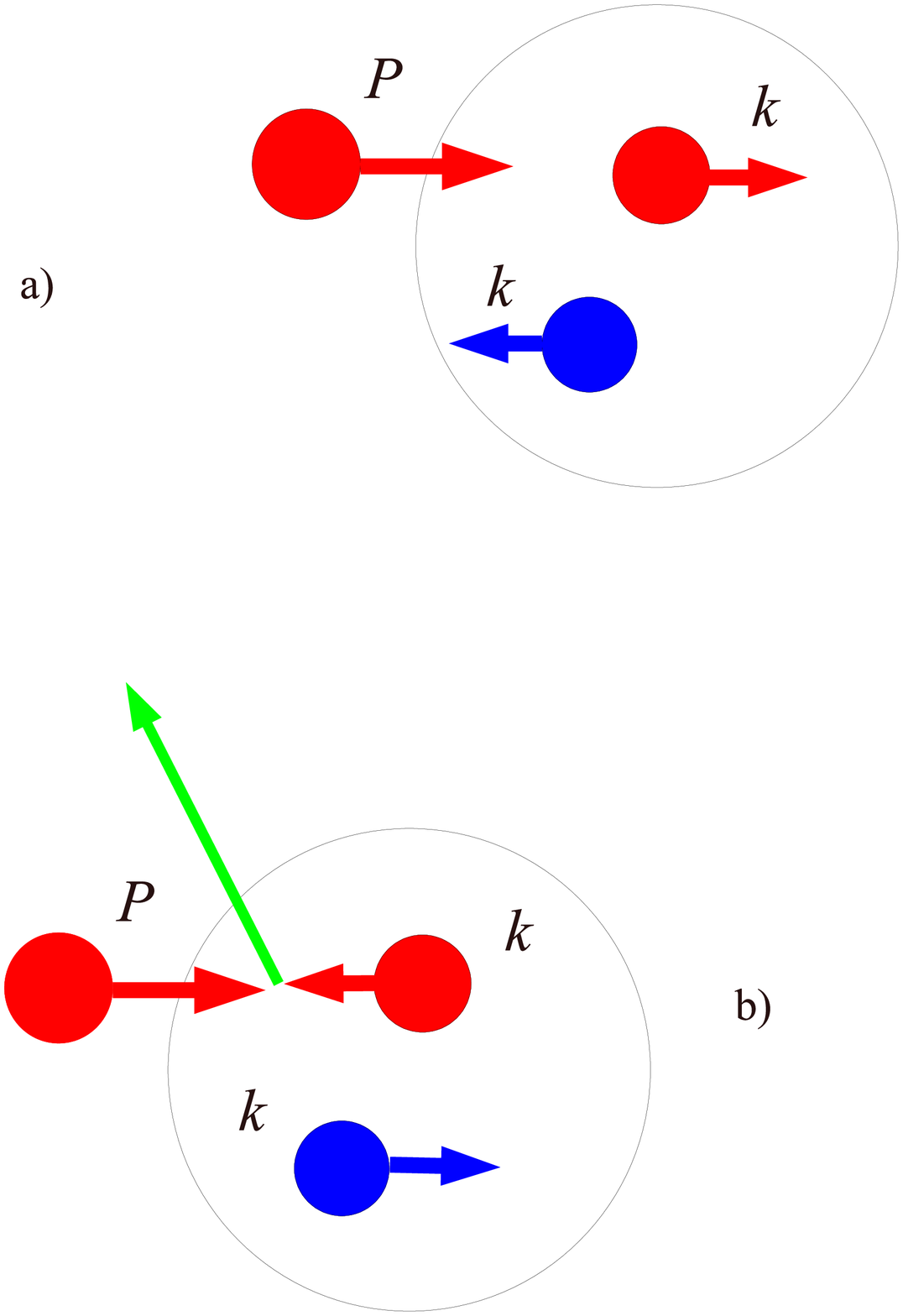} \\
Figure.3 SRC pictures of the backward cumulative particle
production: a)nuclear fragments production; b)mesons and
antiparticles production
\end{center}
The sketch (a) shows $pA \rightarrow p, n, ... + X$ reactions. The
cross sections for these processes will be
$$\sigma \sim n(\bar{k})\cdot \sigma_0,$$ where $n(\bar{k})$ is
a probability to find the nuclear fragment with three-momentum
$\bar{k}$ inside the nucleus. The sketch (b) shows $pA \rightarrow
\pi,K,\bar{p} ... + X$ reactions. The cross sections for these
processes will be
$$\sigma \sim n(\bar{k})\cdot \sigma(NN\rightarrow \pi,K,\bar{p}...+X),$$
where $n(\bar{k})$ is a probability to find nucleon with
three-momentum $\bar{k}$ inside the nucleus.

The second group of models based on possibility of few nucleons to
form a small size hard object named as a flucton. Figure.4 shows
how will produce cumulative particles in the backward direction in
$pA$-collisions.

\begin{center}
\includegraphics[width=70mm,height=50mm]{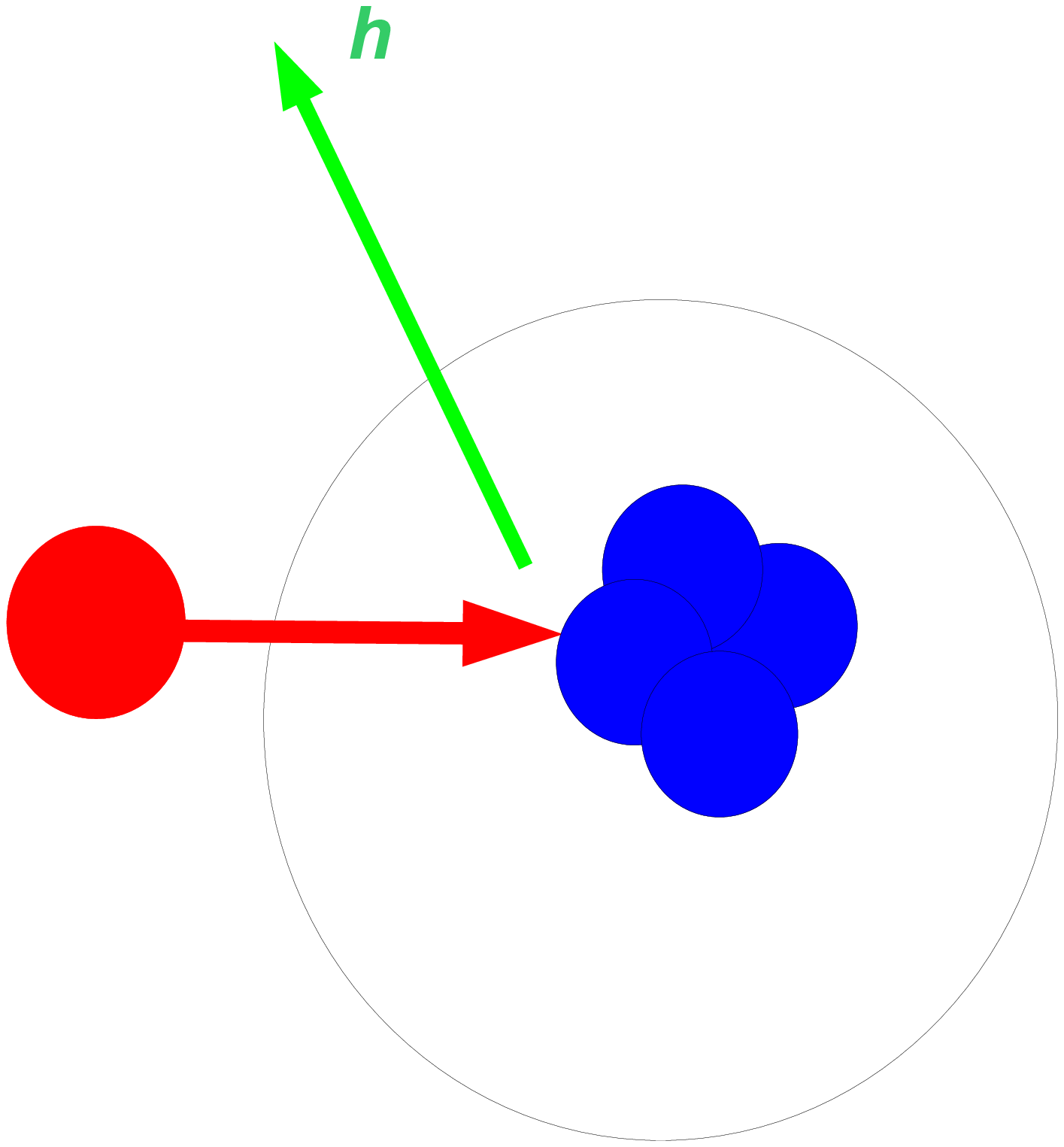} \\
Figure.4 The flucton picture of the backward cumulative particle
production
\end{center}
In these models the cross section for $pA \rightarrow h + X$
reactions will describe by
$$\sigma \sim P_K\cdot G_{h/K}(K),$$ where $P_K$ is a probability
to find the flucton consisting of K nucleons inside the nucleus.
The $G_{h/K}(K)$ gives probability to produce hadron $h$ by this
flucton. Main features of these models are existence of fluctons
and some universal function which describe fragmentation these
fluctons into hadrons.

Studies of the inclusive cumulative particle production in
fragmentation regions of the target or the beam not allow to
separate these two type models. The experiments in low $p_T$
region have huge experimental problems to detect recoil objects
which accompany cumulative particles.

A principally new step in investigation of cumulative phenomena
was made in the experiment E850/EVA [7]. In this experiment for
the first time the effect was studied in a new set up:
semi-exclusive measurements in the maximal $p_T$ region. It is
realised the kinematic presented in Figure.5 (a). In this
experiment two protons (kinematic of the quasielastic
pp-scattering at the angle $90^o$ in the center-of-mass system)
and correlated neutron were detected. It was shown that the
neutron with momentum more then Fermi momentum come mainly from
n-p SRC.

\begin{center}
\includegraphics[width=60mm,height=80mm]{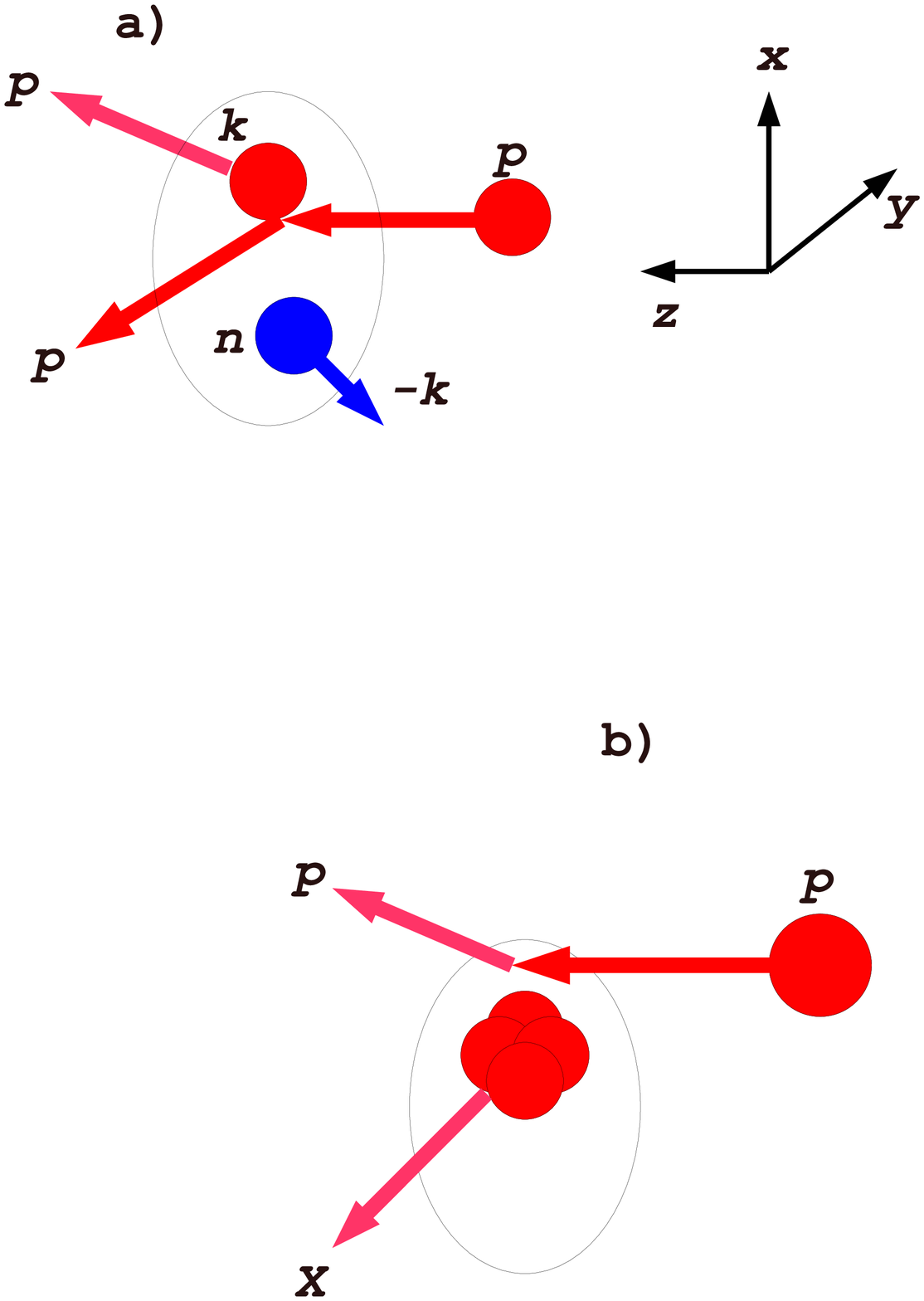} \\
Figure.5 The high $p_T$ cumulative experiments to find SRC (a) or
flucton (b) phenomenons
\end{center}
But there is no possibility to compare this measurement with other
predictions which we can see in Figure.5 (b) as the flucton
mechanism. That is why only [7] data didn't give us answer about
main mechanism for cumulative processes. But it is the first step
which have open for us new way to resolve the cumulative puzzle.
We need more complete investigation in the range of maximal $p_T$
in semi-exclusive (and exclusive) experiment set up for
comprehension of the nature of cumulative processes. It will need
to investigate: \vspace*{2mm}

- average number of baryons accompanied high $p_T$ cumulative
particle production and its $s_{cumulat}$ dependance;\\

- average multiplicity accompanied high $p_T$ cumulative
particle production and its $s_{cumulat}$ dependance;\\

- $s_{cumulat}$ dependence of polarization characteristics
(analyse power, asymmetry and so on), for SRC mechanism will be
scaling repeating effects for free nucleon-nucleon interactions;\\

- coincidence cross sections of high $p_T$ cumulative particle
production with prediction of the "quark counting rules" [9] when
using Stavinsky's variables.

These investigations will give us new important information to
define real mechanisms which respond the processes with $x > 1$.
We could receive proof about SRC or flucton structures inside
nuclei (in the last case we will investigate the some features of
fluctons too). We must stress that up to now we haven't
quantitative predictions for these exclusive(quasi-exclusive)
cumulative characteristics with high $p_T$ from the theoretical
side.

\vspace*{3mm}

{\bf 3.~~High $p_T$ phenomenons}

\vspace*{3mm}

It is symbolically that the new data on cumulative processes [7]
were obtained as a by-product of studies of other nuclear puzzle
named as color transparency of nuclei [8]. It is mean that the
same set up will give us possibility to investigate and may be
resolve a lot of physical problems in the high $p_T$ region. Cross
sections of such processes quickly fall with the energy of beams.
The real possibility for these investigations are limited by beam
energy 20-30 $GeV$. That is why in the nearest time only nuclotron
(Dubna) and U-70 (Protvino) in Russia will be useful for these
studies.

From the point of view of QCD in elastic processes with large
transverse momenta nucleons look like objects comprising of three
point-like constituents. The direct experimental proof of it is a
very nice working of "quark counting rules" [9]. The quark model
makes a proposition that these three observed objects are valence
quarks. The nucleus, from QCD, in "quasielastic" processes with
large transverse momenta looks like a "soft pie" with hard insets
of three valence quarks (nucleons). Therefore, with increasing
primary beam energy the processes of "quasielastic" hard
scattering of hadrons (leptons) on a nuclear target are similar to
elastic scattering on a nucleon target, but a larger number of
scattering centers (nucleons). From QCD point of view the ratio
between cross sections $pA\rightarrow "pp" + X$ and $pp\rightarrow
pp$ processes must smoothly growing to unit. In experiments at the
accelerator AGS(BNL,USA) [8] with proton beams on carbon nuclei
anomalous behavior (of a resonance-type) was observed in a
momentum range of 9.5 GeV/c. This is a so called nuclear color
transparency problem. The nature of this effect is not well
understood yet. A whole physical research program was developed
and proposed [8] to clarify this mystery.

Situation is aggravated by the anomaly of behavior in the same
kinematical region of the elastic proton-proton cross section. It
has no explanation of the strong deviation (about two times) in
the same kinematic region where the "quark counting rules" (the
transverse momentum $>$ 1 GeV/c, the energy of primary beam $>$ 6
GeV) must working very well.

Other item to the portrait of the "crisis" in our understanding of
nucleon-nucleon (nucleon-nuclear) interactions in this region, let
us add that till now the "spin crisis of the 70-se" has not been
solved. There is no understanding of the anomalously strong spin
dependence of the elastic-scattering cross section of protons (at
the angle $90^o$ in the center-of-mass system) for momenta of
protons  8-9 GeV/c. It is real riddle that in many measurements of
cross sections we see that "counting rules" are working very well
(in the pp elastic scattering with maximal $p_T$ too). But naive
quark predictions disagree drastically with the polarization data.
We can propose to investigate reactions $$pp(\bar{p})\rightarrow
BB(\bar{B}) + MM, $$ where B is a baryon and M is a meson. Baryons
must have the large $p_T$. These reactions will give possibility
for more detail studies not only the nucleon quarks structure in
the valence quark dominance region but the spin structure of the
interaction.

Still there is no theory which explains large spin effects in
inclusive processes of the meson and the hyperon production. These
effects do not vanish up to the energies of  hundreds of GeV.
These all show that in the region of nuclotron energies so many
fundamental problems have been accumulated that even a small
number of an additional data can radically help for their
solution.

To carry out all these investigations as cumulative ones in high
$p_T$ region LHE will need to create spacial experimental set up.

\vspace*{5mm}

{\bf 4.~~Summary}

\vspace*{3mm}

In this report we have presented new physical programm which
realization should help to resolve fundamental problems as nature
of the cumulative effect and huge disagreement polarization
phenomenons with predictions naive quark models in high $p_T$
region.

\newpage

\vspace*{3mm}

{\bf References}

\vspace*{5mm}

\begin{tabular}{rp{135mm}}

 [1]&A.M.Baldin "Bulletin of the Lebedev Physics Institute" LPI RAS, {\bf N1}, p.35,
 1971. \\

 [2] &V.S. Stavinsky,PEPAN,{\bf vol.10,issue 5}, p.949, 1979.\\

 [3] &A.A. Baldin, Phys.At.Nucl. {\bf 56(3)}, p.385, 1993.\\

 [4] &V.S. Stavinsky, JINR Rapid Commun. {\bf N18-86}, p.5, 1986.\\

 [5] &Proceedings of the International Symposium "DUBNA
DEUTERON-93", JINR {\bf E2-94-95}, Dubna, 1994; Proceedings of the
3rd International Symposium "DUBNA
DEUTERON-95", JINR {\bf E2-96-100}, Dubna, 1996. \\

 [6] &A.V.Efremov,PEPAN,{\bf vol.13,issue 3},p.613,1982;\\
 &V.V.Burov,V.K.Lukyanov,A.I.Titov,PEPAN,{\bf vol.15,issue 6},p.1249,1984;
\\ &M.I.Strikman,L.L.Frankfurt,PEPAN,{\bf vol.11,issue
3},p.571,1980; \\
&M.I.Strikman,L.L.Frankfurt, Phys.Rep.,{\bf vol.160,issue
5\&6},p.235,1988. \\

 [7] &A.Tang et al.,Phys.Rev.Lett.,{\bf vol.90, N4}, 042301, 2003.\\

 [8] &A.Leksanov tn al.,Phys.Rev.Lett.,{\bf vol.87, N21},212301,2001); \\
&J.Aclander et al.,Phys.Rev\ {\bf C 70}, 015208,2004. \\

 [9] &V.A.Matveev,R.M.Muradyan,A.N.Tavkhelidze,Lett.Nuov.Cim.{\bf vol.7}, p.719,1973 \\
&S.Brodsky,G.Farrar,Phys.Rev.Lett.,{\bf vol.31},p.1153,1973. \\

 [10] &N.I.Golubeva et al.,JINR Report {\bf Ð9-2002-289},Dubna,
 2002. \\

\end{tabular}

\end{document}